## Molecular Dynamics simulations and Kelvin Probe Force microscopy to study of cholesterol-induced electrostatic nanodomains in complex lipid mixtures

E. Drolle[1,2,], W.F.D. Bennett[3], K. Hammond[4], E. Lyman[5], M. Karttunen[6], and Z. Leonenko[1,2,4*]

[1]Department of Biology, University of Waterloo, [2]Waterloo Institute of Nanotechnology, University of Waterloo, Canada, [3]Department of Physics and Astronomy and Department of Chemistry, University of California, Santa Barbara, USA, [4]Department of Physics and Astronomy, University of Waterloo, Canada, [5]Department of Physics and Astronomy, 217 Sharp Lab, Newark, USA, [6]Department of Mathematics and Computer Science & Institute for Complex Molecular Systems, Eindhoven University of Technology, MetaForum, 5600 MB Eindhoven, the Netherlands,
*corresponding author: zleonenk@uwaterloo.ca

### Abstract

The molecular arrangement of lipids and proteins within biomembranes and monolayers gives rise to complex film morphologies as well as regions of distinct electrical surface potential, topographical and electrostatic nanoscale domains. To probe these nanodomains in soft matter is a challenging task both experimentally and theoretically. This work addresses the effects of cholesterol, lipid composition, lipid charge, and lipid phase on the monolayer structure and the electrical surface potential distribution. Atomic Force Microscopy (AFM) was used to resolve topographical nanodomains and Kelvin Probe Force Microscopy (KPFM) to resolve electrical surface potential of these nanodomains in lipid monolayers. Model monolayers composed of dipalmitoylphosphatidylcholine (DPPC), 1,2-dioleoyl-sn-glycero-3-phosphocholine (DOPC), 1-palmitoyl-2-oleoyl-sn-glycero-3-phosphocholine (POPC), 1,2-dioleoyl-sn-glycero-3-[phospho-rac-(3-lysyl(1-glycerol))] (DOPG), sphingomyelin, and cholesterol were studied. It is shown that cholesterol changes nanoscale domain formation, affecting both topography and electrical surface potential. The molecular basis for differences in electrical surface potential was addressed with atomistic molecular dynamics (MD). MD simulations qualitatively match the experimental results, with 100s of mV difference in electrostatic potential between liquid-disordered bilayer ($L_d$, less cholesterol and lower chain order) and a liquid-ordered bilayer ($L_o$, more cholesterol and higher chain order). Importantly, the difference in electrostatic properties between $L_o$ and $L_d$ phases suggests a new mechanism by which membrane composition couples to membrane function.

### Introduction

The plasma membrane of eukaryotic cells is a complex mixture of several hundred lipids and membrane proteins. Lateral variations in lipid composition are thought to be responsible for many biological functions, including initiation of signaling pathways, translocation of small molecules, membrane fusion, and various sorting processes[1, 2]. The functional consequences of "domains" of differing lipid composition likely stem from their ability to bind and sequester membrane proteins. The characterization and definition of these domains has led to a vigorous debate, especially over their characteristic time and length scales[3-8].

Though the plasma membrane contains many different lipid species, the largest fraction is cholesterol, which typically comprises 20-30% by mole. Cholesterol is known to broaden the liquid to gel phase transition, forming an intermediate "liquid-ordered" ($L_o$) phase at sufficiently high concentration[9]. Cholesterol tends to order saturated chains, reducing the area per phospholipid and thickening the membrane while maintaining the fluid phase[10]. Addition of a lower melting temperature lipid introduces a region of $L_o/L_d$ coexistence [11].

Ternary mixtures have been studied extensively over the past decade. The most well-characterized is DPPC/DOPC/Chol, as this mixture forms micron scale domains and can be imaged directly by light microscopy[12, 13]. MD simulations support a model in which cholesterol associates preferentially with saturated acyl chains over unsaturated ones[14-19]. Atomistic MD simulations of the DPPC-DOPC-Chol system have found cholesterol to arrange itself normal to the bilayer surface (upright) at sufficiently high concentrations[14]. More recently, long timescale simulations of the same mixture revealed nanoscale structure within the $L_o$ phase[20]. A coarse-grained MD simulation of DPPC-DOPC-Chol bilayer showed fast partitioning of the mixture into raft-like domains, with the liquid-ordered domains (DPPC) being highly enriched in cholesterol and thicker than the surrounding DOPC regions[16, 21]. Monolayer simulations have shown phase separation similar to the bilayer model, again with cholesterol enrichment in the DPPC phase, at surface pressures from 1-30 mN/m[17]. However, electrostatic effects have been comparatively understudied, and yet offer an obvious mechanism for targeting proteins to specific regions of membranes. Many membrane proteins are directed to the membrane by nonspecific electrostatic interactions[22].

Direct imaging of lipid bilayers on the nanoscale is difficult due to their soft nature — they are held together by hydrophobic interactions within a 2 molecule thick membrane (~5 nm) solvated by two chemically diverse water interfaces. Atomic force microscopy (AFM) is one of the few methods which provides direct, nanoscale information on membrane organization[23-27]. Of particular interest to the present work is Kelvin Probe Force Microscopy (KPFM), a high-resolution, sensitive method for direct measurement of surface potential, recently used to map surface potentials in model lipid systems[28, 29]. For example, the presence of cholesterol causes electrostatic domain formation in a pulmonary surfactant lipid mixture containing DPPC and anionic lipids[28, 29].

In the following, KPFM is used to characterize cholesterol induced electrostatic domains in model systems. Combination of AFM with KPFM allows for measuring of both topographical and electrostatic domains. Atomistic MD simulations used to model these domains, support the experimental results, demonstrate that the results extend to bilayer systems, and provide insight into the mechanistic details responsible for the electrostatic domains.

## Materials and Methods

### Lipid solution preparation

Dipalmitoylphosphatidylcholine (DPPC), 1,2-dioleoyl-*sn*-glycero-3-phosphocholine (DOPC), 1,2-dioleoyl-*sn*-glycero-3-phospho-(1'-*rac*-glycerol) (sodium salt) (DOPG), and cholesterol were purchased from Avanti Polar Lipids (Alabaster, AL) in powder form. All other chemicals used, including chloroform and ethanol, were of reagent grade. Stock lipids were dissolved in chloroform at a concentration of 1 mg/mL and combined to form the mixtures shown in Table 1. These solutions were then used for monolayer deposition with the Langmuir-Blodgett trough.

### Supported lipid monolayers preparation on the Langmuir-Blodgett trough.

For monolayer preparation, phospholipids were deposited on fresh cleaved mica (Asheville-Schoonmaker Mica Co., Newport News, VA) using Langmuir-Blodgett deposition. In this method, phospholipid mixtures in chloroform are spread onto the surface of liquid subphase. The molecules orient themselves into a monolayer at the air/water interface, and are then compressed via moveable barrier arms at a rate of 10 cm$^2$/min to a pressure of 35 mN/m. At this pressure, the moveable dipper arm of the trough arm passes the mica substrate through the monolayer, utilizing vertical deposition in order to create a supported monolayer, while the barrier arms maintain the pressure of 35 mN/m. Samples were then allowed to dry in a dessicator for a minimum of 24 hours and affixed on a conductive plate for AFM/KPFM imaging.

### Atomic force microscopy (AFM) and frequency modulated-Kelvin probe force microscopy (FM-KPFM).

Topography imaging (AFM) of the supported monolayers was performed simultaneously with electrical surface potential mapping (FM-KPFM) in air using intermittent contact mode on the AIST-NT SmartSPM, with a Micromasch chromium-gold coated cantilever (HQ:NSC14 / Cr-Au) with a resonance frequency of 160 kHz and a spring constant of 5.0 N/m. The use of FM-KPFM allows for higher resolution than typical KPFM techniques and simultaneous AFM-KPFM imaging.

**Image processing and analysis.**

Image processing was carried out using AIST Image Analysis and Processing software. Images were levelled using plane levelling and lines were corrected by fitting the lines and removing scars (any image artifacts present). Statistics were obtained using roughness analysis made available by the AIST Image Analysis and Processing software, as well as cross sectional analysis of multiple samples and scans for each combination. For analysis of FM-KPFM images, raw data was used; the images were not processed with any filters prior to the analysis in order to ensure that the potential measurements were accurate. All quantitative results are presented as mean ± standard error and differences in the presence of cholesterol are statistically significant based on a *t*-test statistical analysis with confidence level of at least 90%, unless otherwise indicated.

**Molecular Dynamics Simulations**

Atomistic MD simulations were conducted using GROMACS v4 [30]. The Stockholm lipid parameters [31] that are based on the AMBER force field were used for DOPC, DPPC, DOPG and cholesterol. Detailed comparisons between different lipid models indicate that currently CHARMM36 [32] appears to be the best overall force field [33, 34], but when cholesterol is present, Stockholm lipid parameters work well and, importantly for this study, the effects of cholesterol on PCs are better reproduced[35]. Water was modelled with the TIP3P model [36]. Systems were constructed using the Insane method [37] and the Martini force field [38]. After a 50 ns equilibration with Martini, the Backwards method [37] was used to convert the coarse-grained solvated bilayer to atomistic representation and the simulation was continued. A switch function was used for Lennard-Jones interactions from 1.4 nm to 1.5 nm, with a dispersion correction. Short-range electrostatics were truncated at 1.0 nm, with the particle mesh Ewald algorithm[39] for long-range interactions. The Nosé-Hoover thermostat[39, 40] was used for maintaining constant temperature with a 0.5 ps coupling constant. Semi-isotropic pressure coupling was used to maintain the pressure at 1 bar in the plane of the membrane and normal to the membrane plane. The Parrinello-Rahman barostat[41] was used with a compressibility of $4.5 \times 10^{-5}$ $bar^{-1}$ and a 10 ps coupling constant.

We used small lipid bilayers as mimics for the bulk phases of the AFM experiments. The gel-phase bilayer was 72 DPPC lipids (only the last 50 ns were used for analysis; after it reached equilibrium in the gel-phase). The DOPC:DPPC:CHOL ratio for the $L_d$ (42:20:6) and $L_o$ (10:38:20) phase was taken from the ternary phase diagram. Simulations were also performed with DOPG lipids in compositions of 44:20:6 and 10:38:20 DOPG:DPPC:CHOL in order to investigate the influence of headgroup charge on the potential, with $Na^+$ to neutralize the charge of the PG head groups.

The electrostatic potential ($\Psi$) across the bilayer was determined by integrating the charge distribution ($\rho$) twice according to Poisson's equation:

$$\frac{d^2\psi(z)}{dz^2} = -\frac{\rho(z)}{\varepsilon_0}$$

Boundary conditions were chosen such that the electric field and potential are zero at the center of the bilayer[42]. After computing the potential, the curves were shifted to be equal to zero in bulk water, as we are interested in the difference in potential between water the membrane center. For the electrostatic potential decompositions, the charge distribution is calculated for a particular chemical group (e.g., water) which is then numerically integrated to obtain the potential[43].

## Results

### Ternary DPPC:DOPC:CHOL systems

In order to study the effect of cholesterol on multi-lipid systems as well as the formation or abolishment of lipid domains, we looked at samples with and without cholesterol. A schematic of two-component lipid monolayer system is given in Figure 1. The image also depicts how difference in height and difference in electrical surface potential are measured. Differences in height ($\Delta$h) are imaged via AFM topography analysis of the sample, and differences in electrical surface potential ($\Delta$V, equivalent to Vtopo(high) – Vtopo(low)) of the same samples (though not visible in this image) are measured by FM-KPFM electrical surface potential mapping. When cholesterol is present in the mixture it is found in both phases (saturated and unsaturated lipid phases), and thus the properties ($\Delta$h and ($\Delta$V) of the domains change.

**Figure 1: Schematic of Monolayer Arrangement in a Two-Component Lipid System.** The hydrophilic head groups of the lipids interact with the hydrophilic surface of the mica substrate, resulting in the tail groups facing upwards into the air. The difference in height ($\Delta$h) is extracted from AFM topography and difference in electrical surface potential ($\Delta$V, equivalent to Vtopo(high) – Vtopo(low)) is determined using KPFM.

Figure 2A presents results for a DPPC-DOPC mixture with and without cholesterol. In the absence of cholesterol, domains are observed both in AFM and KPFM. The difference in height between these domains ($\Delta$h, visualized in Figure 1) averaged to $1.22 \pm 0.03$ nm. It is likely that higher domains correspond to DPPC molecules, as in their gel phase, the tail groups are more ordered and thus are slightly thicker than the disordered tail groups of fluid phase (also known as liquid disordered, $L_d$, phase) DOPC[44]. These higher domains were much larger in size than the lower domains, reaching lateral dimensions of up to 570 nm in length (i.e. X) and 775 nm in width (i.e. Y); the lower domains were less prevalent in the monolayer, with average X dimensions of 150 nm and Y dimensions of up to 350 nm. The domains also have distinct electrical surface potentials. We analyzed the difference in the electrical surface potential ($\Delta$V) between the higher and lower domains (visualized in Figure 1). The average $\Delta$V of the electrostatic domains was determined to be 41.6 ± 5.39 mV (Figure 2B).

With the addition of cholesterol to the DPPC-DOPC system, smaller topographical domains on the monolayer samples (Figure 2C) with an average $\Delta$h of the domains of 0.97 ± 0.06 nm were observed. This average $\Delta$h is slightly lower than that observed in the DOPC-DPPC system without cholesterol (1.22 nm). This is consistent with the idea that cholesterol causes slight disorder in the already ordered tail groups of gel phase lipid molecules such as DPPC by creating an intermediate $L_o$ phase[16, 44, 45]. The idea that cholesterol molecules are clustered in higher concentration near the DPPC molecules than the DOPC molecules agrees with the idea that cholesterol prefers to interact with saturated rather than unsaturated lipids[44]. It also agrees with many simulation studies that show cholesterol molecules organized in higher concentration with DPPC molecules rather than the DOPC molecules[15].

The corresponding surface potential image (Figure 2D) shows distinct electrostatic domains where higher topographical domains correspond to areas of higher electrical surface potential. These electrostatic domains had an average $\Delta$V of 67.25 ± 7.03 mV. This difference is larger than that observed in the DOPC-DPPC-Chol system, which had an average $\Delta$V of 41.6 mV, supporting the idea that cholesterol has a measurable effect on the V of the mixed lipid system.

Figure 3 shows snapshots of the model bilayer systems that were simulated. To mimic the DPPC-DOPC, gel-liquid phase separated monolayer system, a DPPC bilayer simulation was run at 298 K for 300 ns, equilibrating to a gel-phase. The straight and aligned tails in the gel phase are clearly visible in Figure 3. A DOPC bilayer also at 298 K was run for 100 ns, equilibrating to a fluid phase. The compositions of the ternary mixtures in the presence of cholesterol are described in Methods. The Na+ ions, rendered as green spheres in Fig. 3, bind to the negatively charged DOPC head groups.

Figure 4 compares the electrostatic potential and mass density profile for the gel DPPC and liquid DOPC bilayer. For the gel bilayer, due to tight packing and slow dynamics, two distinct density peaks are observed for the gel-DPPC bilayer head group region, whereas for the liquid DOPC there is a smooth distribution for the head groups. Qualitatively the simulations match the AFM results; the gel bilayer is thicker than the liquid DOPC phase. Simulations show a sharp drop in density near the bilayer center, because the two gel leaflet tails cannot interdigitate. This is, however, somewhat complicated by the tilt of the DPPC hydrocarbon chains in the gel phase, Fig. 3.

The electrostatic profile (Fig 4C) is shifted to longer distances for the gel bilayer, due to the increased bilayer thickness. The potential is higher for the gel bilayer, until right at the bilayer center, where the density drops, and the potential for the liquid DOPC becomes slightly larger. This is different than the KPFM results on the DPPC-DOPC phase separated monolayer systems, where the gel DPPC system had a larger potential compared to DOPC. We speculate that the discrepancy could be due to differences between a bilayer and monolayer system, given that the coupling of the gel leaflets is not present in a monolayer. The gel DPPC electrostatic potential is larger than DOPC, until right at the low density bilayer center. Despite similar total electrostatic potentials, the components are quite different, with ~1 V difference (although we note that these contributions are not 'felt' by the system due to explicit screening of the charges, to yield the total electrostatic potential). Water at the interface is ordered resulting in a positive dipole potential that offsets the large negative dipole potential at the bilayer center from the negative phosphate group's position relative to the positive choline group.

Fig 5 shows electrostatic and density profiles across cholesterol mixtures, a 10:38:20 (DOPC:DPPC:CHOL) for the $L_o$ phase, and 42:20:6 at 298 K for the $L_d$ phase, both at 298 K. Cholesterol is known to order saturated lipids, resulting in the $L_o$ phase, which has structural order similar to the gel phase, but is still liquid with relatively fast lateral diffusion of lipids compared to the gel phase. Density profiles show that the bilayer is thicker for the $L_o$ bilayer, but the $L_o$ bilayer's density is more similar to the $L_d$ bilayer than the gel system, without distinct phosphate and choline peaks. Due to the thicker density, the potential profile is shifted out towards water for the $L_o$ and gel phase. There is also a small trough in the electrostatic potential near the center of the $L_O$ similar to the gel phase systems. The electrostatic potential at the center of the $L_o$ phase (1.11 V) is higher compared to the $L_d$ phase (1.03 V), by a difference of 80 mV. Considering the limitations of fixed charge models for bilayer electrostatics, this difference is very similar to the measurements of differences in electrostatic surface potential of 35 to 70 mV obtained by KPFM (Table 2). Because these are bilayers compared to supported monolayers for the KPFM results, we would not expect perfect agreement. Of note, the electrostatic contributions from water and the lipid bilayer are very similar for the $L_o$ and $L_d$ bilayers, as opposed to the gel bilayer (Fig. 4).

Comparison between the monolayer and bilayer systems is not, however, straightforward[46]. Since the monolayer DPPC-DOPC-Chol systems were prepared at a relatively high surface pressure of 35 mN/m and low cholesterol concentration corresponding to the so-called α-region[46, 47], it is likely that they are in the gel, or liquid-condensed phase. This does not directly correspond to the $L_o$ phase. Although beyond the scope of the current study, direct comparison (to $L_o$) would require surface pressures around or below 15 mN/m[47]. We used the compression of 35 to mimic physiological compressions reported for the membrane[48-50]. We can, however, make the following conclusions: consistent with experiments, the lipid tail in the gel phase of the pure DPPC display the characteristic tilt[51] which is absent in the $L_o$ phase mixtures when cholesterol is interacting with them; addition of cholesterol or DOPC leads to straightening of the DPPC hydrocarbon chains even in the gel phase[52].

**Negatively charged lipid system**

DPPC, and DOPC, have a neutral net charge, so differences in surface potentials arise from the dipole moments of the lipids and their ordering. However, it is of interest to see if cholesterol influences the surface potential of a sample that contains a lipid with a charged head group. To this end, samples of DPPC (charge neutral, zwitterionic head group) and DOPG (net negatively charged head group) were examined. AFM and FM-KPFM images of this system are shown in Figure 6.

Figure 6A shows the topography for the pure DPPC-DOPG system. Distinct differences in topography are observed that likely arise due to the phase separation of gel phase DPPC and fluid phase DOPG. We assume that higher domains are saturated with gel phase DPPC and lower domains are saturated with fluid phase DOPG. The average difference in height, Δh, between these domains was determined to be 0.80 nm ± 0.04 nm. The corresponding KPFM image shows differences in potential (Figure 6B), average ΔV is -138.67 ± 6.79, Table 2.

In the DPPC-DOPG-Chol image larger domains are observed than in the cholesterol-free sample (Figure 6C). These surface features have an average Δh of 1.18 ± 0.06 nm. There also appear to be areas of the sample consistent with multilayer formation. This feature is normally seen at higher pressures, when the lateral pressure exerted on the lipid monolayer causes the monolayer to buckle and fold, alleviating the high pressure.

The corresponding FM-KPFM image shows electrostatic domains in areas that correlate with topographical domains (Figure 6D). Though cholesterol's effect on the ΔV – an average of -117.55 ± 8.65 mV for this sample – is only slightly smaller than the -138.67 mV for the DPPC-DOPG sample, Table 2, it does show that cholesterol influences surface potential in lipid systems containing charged head groups.

In order to make comparison with simulations we need to compare the electrical surface potential between DPPC/Chol phase and DOPG/Chol phase, assuming that domains we observed are result of DPPC – DOPG demixing and that cholesterol is present in each lipid phase. The electrostatic profile for the PG bilayers are slightly different than the other systems, with a small negative potential near the water interface (Fig. 5B), and a significantly lower potential at the center of the bilayer. For pure DPPC (gel) and DOPG (fluid) bilayers, the electrostatic potential difference is 40 mV. This result is different from experiments, most likely due to effect of counter-ions that were accounted for in bilayer system, in MD simulations. In contrast, KPFM experiments are done on supported lipid monolayer, which was deposited on mica by Langmuir-Blodgett deposition from the water – air interface, therefore there are no counter-ions present in this system.

Comparing the electrostatic potential between the PG-$L_D$ bilayer (0.89 V) and the PG-$L_O$ phase (0.96 V), we find a difference of 70 mV. Because the PG lipid head group has a net charge, and $Na^+$ ions were present to keep the system neutral, the electrostatic contributions from these groups are large. The contribution from water is similar for the PG-$L_o$ and PG-$L_d$ bilayers (~11 V), which is considerably larger than the $L_o$ (~4 V), indicating strong ordering of water at the interface of the charged lipid system. Despite the large negative surface charges from PG, there is only a small negative total electrostatic potential at the interface, due to the accumulation of $Na^+$ on the surface to shield the head group charges; very recent studies indicate the need to improve the ionic force fields used in lipid simulations[53]. Similar concerns regarding the lipid phases of the monolayers and bilayers as discussed in connection with the mixed DPPC-DOPC-Chol exist. When PG lipids are present, uncertainties in the ion-PG interactions contribute the quantitative discrepancies[53].

The results of MD simulations are summarized in Table 3. Despite the discrepancies, both results are valuable in assessment of complex matter of understanding of electrical surface potential in lipid monolayers and bilayers.

## Discussion

Cholesterol affects both the height and the difference in electrical surface potential in all lipid systems studied here, Table 2 and Table 3. Cholesterol orders saturated hydrocarbon chains, which changes the distribution of charge and density normal to the bilayer. In addition, cholesterol drives lateral domain formation, which manifests as variations in the electrical surface potential as observed by FM-KPFM[54]. Furthermore, because of the cholesterol's orientation in the monolayer, and its interaction with both the head group and the ester linkage between the head group and fatty acid tails, it has the ability to affect both the dipole moment of other molecules as well as the tilt.

Using MD simulations, it is possible to measure the effect that cholesterol has on a single molecule's electrical surface potential or on a single component lipid system's membrane dipole potential, as seen in Starke-Peterkovic et al.'s work[54]; however, when it comes to multiple molecules, it becomes a much more complex issue. This is the case with domains within the membranes, which could contain thousands of interacting molecules. This work presents a first approach where experimental AFM and KPFM imaging were combined with MD simulations to address both topographical and electrostatic nanodomains in the multicomponent lipid systems. Extending this approach to more complex mixtures should help determine whether such electrostatic domains are a feature of the plasma membrane of intact cells.

**Conclusion**

Overall, these results show that cholesterol has a measureable and significant effect on electrical surface potential in all the systems studied. The experimental results obtained for monolayers are compared to simulations of lipid bilayers of the same composition. This further supports the idea that the presence of cholesterol in the membrane, and in particular its fundamental effect on the electrical surface potential of the membrane, may be largely involved in the membranes interactions with polar or charged biomolecules in nature. Although both monolayer and bilayer systems are videly used to model bio-membranes, we show that elelctrical surface potential can be significantly different in monolayer and bilayer systems due to the presence or absence of counter-ions.

**Acknowledgements**

We acknowledge funding from Natural Sciences and Engineering Research Council (NSERC) of Canada (ZL,MK), Canadian Foundation for Innovation (CFI) and Ontario Research Fund (ORF) (ZL). We thank NSERC's Sir Frederick Banting Fellowship Program (WFDB), NSERC Canada Graduate Scholarship and Waterloo Institute for Nanotechnology (WIN) Graduate Student Fellowship (ED), and NSERC Undergraduate Research Award (KH). Compute Canada and SharcNet provided the computational resources.

| Lipids | Ratio (w/w) |
|---|---|
| DPPC-DOPC | 607:393 |
| DPPC-DOPC-Chol | 560:346:94 |
| DPPC-DOPG | 50:50 |
| DPPC-DOPG-Chol | 40:40:20 |

**Table 1: Lipid Mixtures Studied and Their Respective Lipid Ratios.** The 4 samples analyzed in this study. The corresponding lipid ratios (based on weight) are given.

| | Average Difference in Height $\Delta h$ (nm) | Average Difference in Electrical Surface Potential $\Delta V$ (mV) |
|---|---|---|
| **DPPC-DOPC** | $1.22 \pm 0.03$ | $41.60 \pm 5.39$ |
| **DPPC-DOPC-Chol** | $0.97 \pm 0.06$ | $67.25 \pm 7.03$ |
| *Effect of Cholesterol = Decrease in $\Delta h$, Increase in $\Delta V$* | | |
| **DPPC-DOPG** | $0.80 \pm 0.04$ | $138.67 \pm 6.79$ |
| **DPPC-DOPG-Chol** | $1.18 \pm 0.06$ | $117.55 \pm 8.65$ |
| *Effect of Cholesterol = Increase in $\Delta h$, Decrease in $\Delta V$* | | |

**Table 2: Summary of the Effect of Cholesterol on All Systems Studied.** Average difference in height (in nm) and average difference in electrical surface potential (in mV) are shown for both the cholesterol-free sample and the cholesterol-present sample. All reported effects of cholesterol, unless otherwise indicated, are significant with a confidence level of at least 90% (based on t-test statistical analysis).

| Lipids composition | Calculated potential (V) |
|---|---|
| DOPC–DPPC–Chol ($L_d$) | -1.03 |
| DOPC–DPPC–Chol ($L_o$) | -1.11 |
| DOPG–DPPC–Chol ($L_d$) | -0.89 |
| DOPG–DPPC–Chol ($L_o$) | -0.96 |
| DPPC (gel) | -0.89 |
| DOPG | -0.85 |
| DOPC | -1.03 |

**Table 3.** Summary of results of MD simulations for electrical potential calculated for lipid bilayers. This table summarized the quantitative information illustrated in Fig. 4, 5 and 7

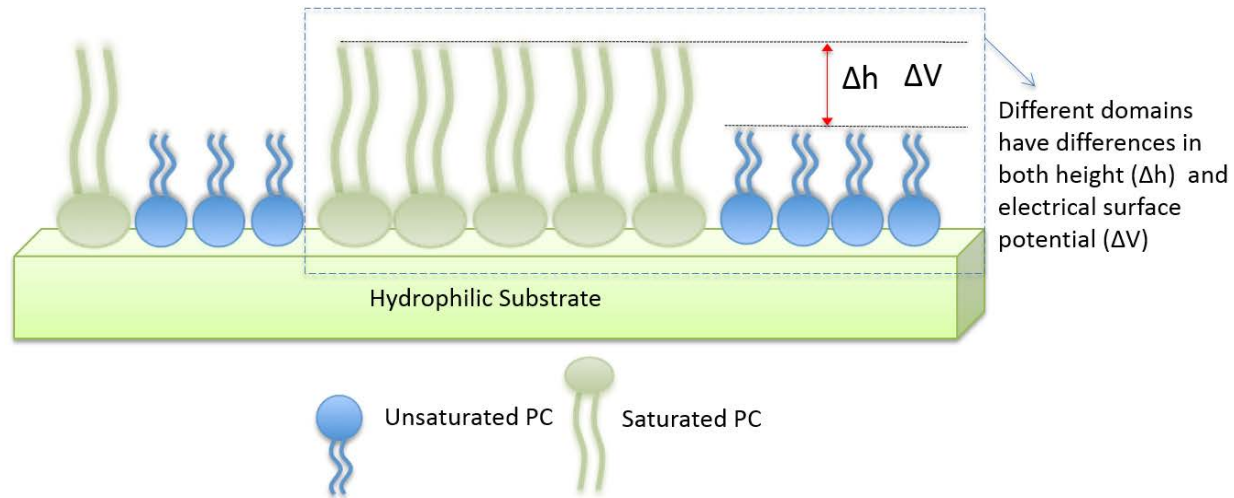

**Figure 2: Schematic of Monolayer Arrangement in a Two-Component Lipid System.** The hydrophilic head groups of the lipids interact with the hydrophilic surface of the mica substrate, resulting in the tail groups facing upwards into the air. The difference in height (Δh) is extracted from AFM topography and difference in electrical surface potential (ΔV, equivalent to Vtopo(high) – Vtopo(low)) is determined using KPFM.

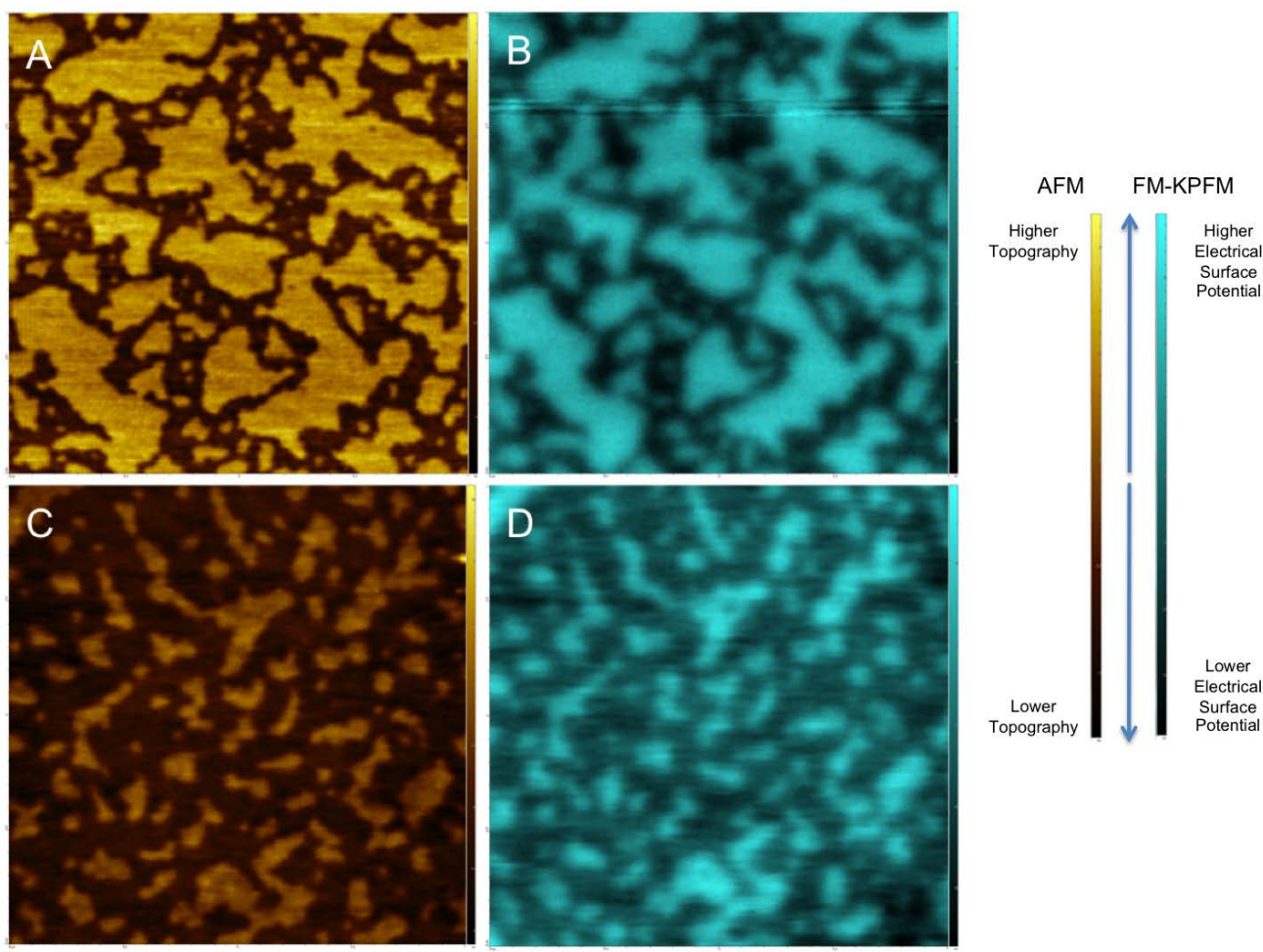

**Figure 3: AFM (in gold) and corresponding FM-KPFM (in blue) illustrating the effect of cholesterol on both the topography and electrical surface potential of a lipid mixture of DPPC and DOPC.** AFM images (A) and FM-KPFM images (B) of a DPPC-DOPC mixture (at a ratio of 607:393) are compared to topography captured by AFM and electrical surface potential captured by KPFM of a DPPC-DOPC-Cholesterol sample (at a ratio of 560:346:94, images C and D respectively). AFM and KPFM images were scanned in air in ambient conditions. These images are 2 μm by 2 μm; brighter regions correspond to higher topography/electrical surface potential, with darker regions corresponding to regions of lower topography/electrical surface potential, as shown in the labelled scale bars to the right.

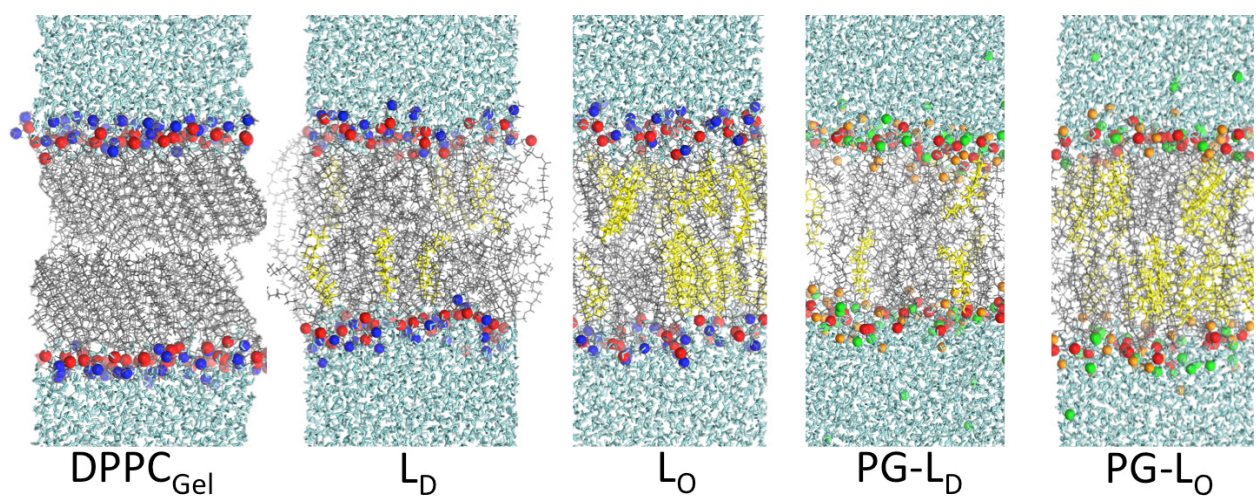

**Figure 3.** Snapshots of the different simulated systems. Cholesterol: yellow, water: cyan, lipid tail: grey, phosphorous: red, nitrogen: blue, oxygen on PG head group: orange, and Na+: green. $L_o$ and $L_d$ denote liquid-ordered and liquid-disordered, respectively.

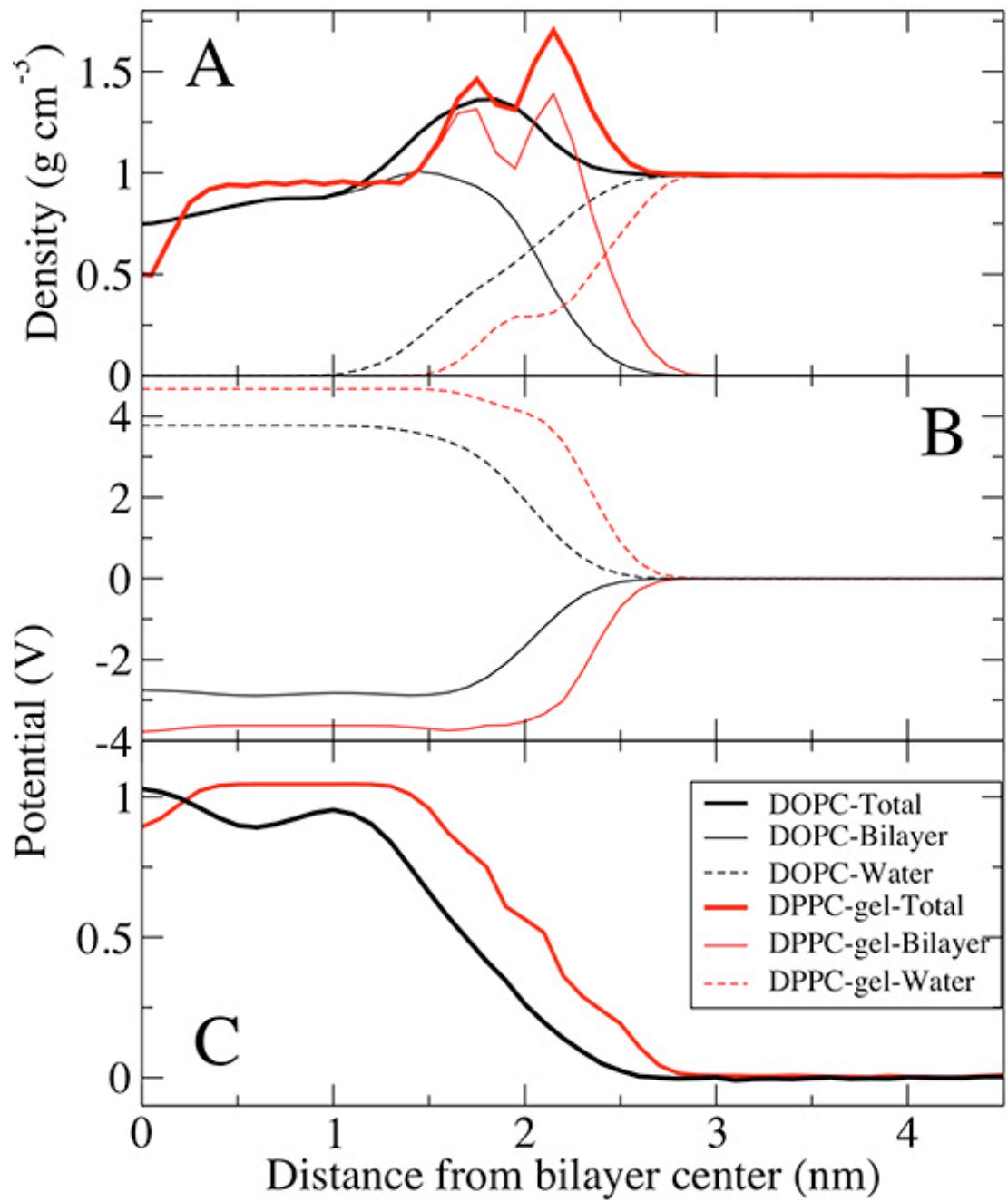

**Figure 4.** Density profiles and electrostatic potentials across one leaflet of the gel DPPC (pure) bilayer and the liquid DOPC (pure) bilayer. A. Density profiles of one leaflet of a fluid phase DOPC bilayer (black) and DPPC gel phase bilayer (red). Total density and contributions of water and lipids are shown separately. B. Contributions to the electrostatic potential from the lipids and water for both systems (see Methods). C. Total electrostatic potential in both systems. The electrostatic potentials were shifted to zero in bulk water for comparison.

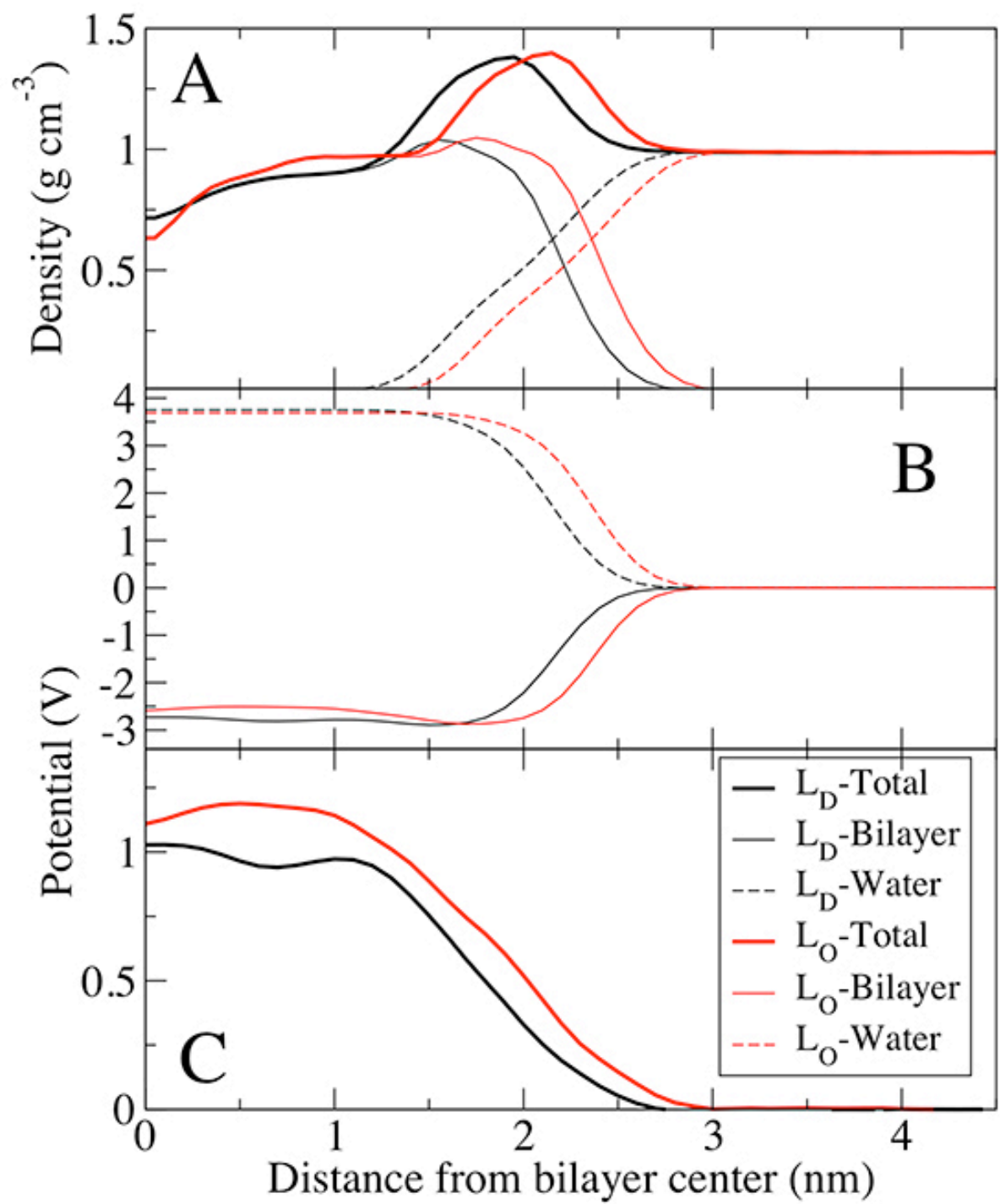

**Figure 5.** Density profiles and electrostatic potentials across one leaflet of the ternary mixtures of DPPC/DOPC/Chol bilayers. A. Density profiles of one leaflet of the $L_D$ bilayer (black) and $L_O$ bilayer (red). Total density as well as contributions of water and lipids are shown separately. B. Contributions to the electrostatic potential from the lipids and water for both systems (see Methods). C. Total electrostatic potential in both systems. The electrostatic potentials were shifted to zero in bulk water for comparison.

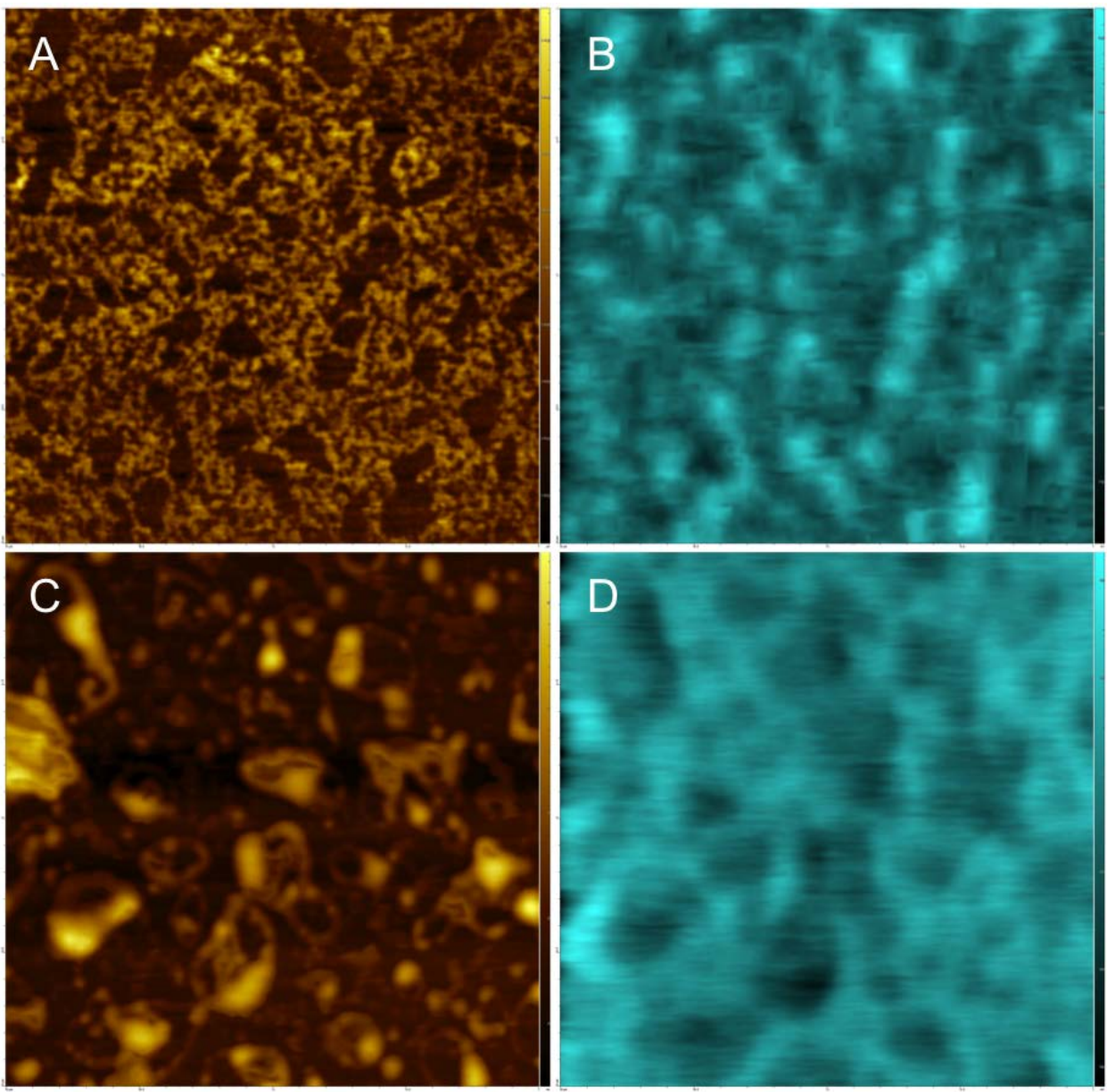

**Figure 6: AFM and corresponding FM-KPFM illustrating the effect of cholesterol on both the topography and electrical surface potential of a lipid mixture of DPPC and negatively charged DOPG.** AFM images (A) and FM-KPFM images (B) of a DPPC-DOPG mixture (50:50 w/w) are compared to topography captured by AFM and electrical surface potential captured by FM-KPFM of a DPPC-DOPG-Chol (40:40:20 w/w) sample (images C and D respectively). AFM and FM-KPFM images were scanned in air in ambient conditions. These images are 2 μm by 2 μm.

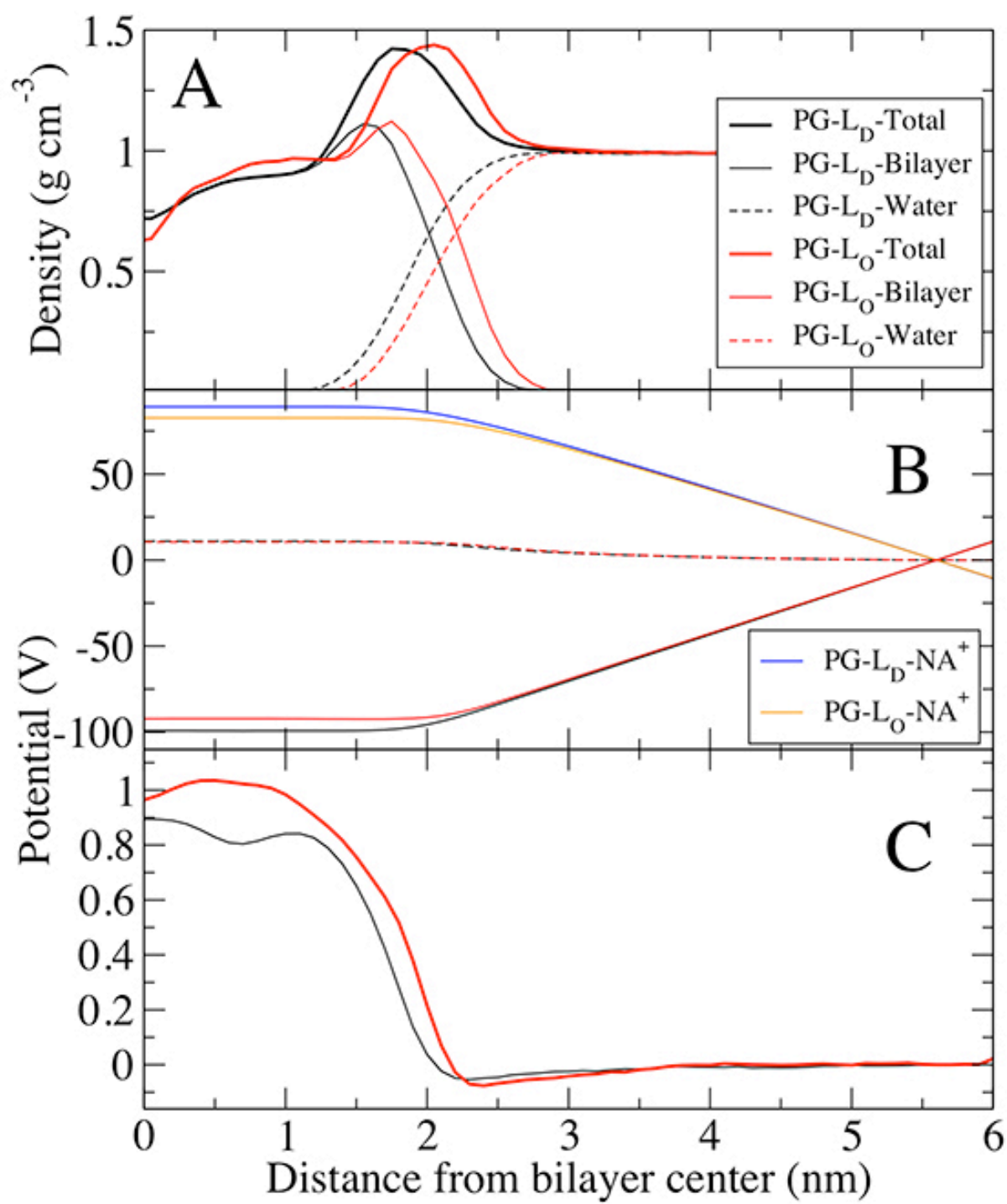

**Figure 7** Density profiles and electrostatic potentials across one leaflet of the ternary mixtures of DPPG/DOPG/Chol bilayers. A. Density profiles of one leaflet of the $L_D$ bilayer (black) and $L_O$ bilayer (red). Total density as well as contributions of water and lipids are shown separately. B. Contributions to the electrostatic potential from the lipids and water for both systems (see Methods). C. Total electrostatic potential in both systems. The electrostatic potentials were shifted to zero in bulk water for comparison.